\documentclass[12pt,reqno]{amsart}
\usepackage{amsmath, amsthm, amssymb}
\usepackage{enumerate}
\usepackage{mathrsfs}
\usepackage{color}
\usepackage[english]{babel}
\usepackage{graphicx}
\usepackage{hyperref}
\usepackage{subcaption}
\usepackage{float}

\newcommand{\sech}{\, \mathrm{sech} }

\usepackage{amssymb}
\usepackage{epsfig}
\usepackage{extarrows}
\usepackage{amsfonts}
\usepackage{graphicx}
\usepackage{color}
\hypersetup{
    colorlinks=true,
    linkcolor=blue,
    filecolor=magenta,      
    urlcolor=blue,
}

\newcommand{\R}{\mathbb{R}}

\newtheorem{remark}{Remark}

\begin{document}
	
	\title[Solitons on the rarefactive wave background]{\bf Solitons on the rarefactive wave background \\ via the Darboux transformation}

	\author{Ana Mucalica}
	\address[A. Mucalica]{Department of Mathematics and Statistics, McMaster University,
		Hamilton, Ontario, Canada, L8S 4K1}
	\email{mucalica@mcmaster.ca} 
	
	\author{Dmitry E. Pelinovsky}
	\address[D. Pelinovsky]{Department of Mathematics and Statistics, McMaster University, Hamilton, Ontario, Canada, L8S 4K1}
	\email{dmpeli@math.mcmaster.ca}

	\begin{abstract}
Rarefactive waves and dispersive shock waves are generated from the step-like initial data 
in many nonlinear evolution equations including the classical example of the Korteweg-de Vries (KdV) equation. When a solitary wave is injected 
on the step-like initial data, it is either transmitted over the background or trapped in the rarefactive wave. We show that the transmitted soliton can be obtained by using the Darboux transformation 
for the KdV equation. On the other hand, no trapped soliton can be obtained 
by using the Darboux transformation and we show with numerical 
simulations that the trapped soliton disappears in the long-time dynamics of the rarefactive wave. 
	\end{abstract}   


\date{\today}
\maketitle

\section{Introduction}

The Korteweg--de Vries (KdV) equation is a classical model for long surface gravity waves of small amplitude propagating unidirectionally over shallow water of uniform depth \cite{Ablowitz}. The normalized version of the KdV equation takes the form 
\begin{equation}\label{1}
    u_t+6uu_x+u_{xxx}=0,
\end{equation}
where $t$ is the evolution time, $x$ is the spatial coordinate for the wave propagation, and $u$ is the fluid velocity. The KdV equation has predominantly been studied on spatial domains with either decaying or periodic boundary values. However due to many applications, e.g. the tidal bores or the earthquake-generated waves \cite{El}, it is also relevant to consider the initial-value problem with the step-like boundary conditions:
\begin{equation}
\label{bc}
    \lim_{x\rightarrow -\infty} u(t,x) = 0, \quad 
    \lim_{x\rightarrow +\infty} u(t,x) = c^2,
\end{equation}
where $c^2>0$ is a constant. 

Evolution of the step-like data results in the appearance of a rarefactive wave (RW) if $t$ advances to positive times or a dispersive shock wave (DSW) if $t$ advances to negative times \cite{ElHoffer}. In what follows, we will consider the initial-value problem (\ref{bc}) for the RW in positive time $t > 0$ since the analysis for negative time $t<0$ is similar \cite{Maiden}.

Since the interaction of waves with a mean flow is an important and well-established problem of fluid mechanics, it has been an active area of research. An excellent account on the hydrodynamics of optical soliton tunneling is given in \cite{Sprenger}, where a localized, depression wave (known as the dark soliton) of the one-dimensional defocusing nonlinear Schrödinger (NLS) equation interacted with either RW or DSW. Another example of the interaction between localised solitary waves with large-scale, time-varying dispersive mean flows was studied in \cite{Sande} in the context of the modified KdV (MKdV) equation. A dual problem was the interaction of a linear wavepacket (modulated waves) with the step-like initial data \cite{Congy}. Both transmission and trapping conditions of a small-amplitude, linear, dispersive  wave propagating through an expansion (RW) or a DSW (undular bore) was explained by using  modulation equations. 

Focusing versions of the same problem were also considered in the cubic NLS equation \cite{Biondini,BM} and in the modified KdV equation \cite{Grava}. Due 
to modulational instability, various solitary waves, breathers, and rogue waves were generated from the step-like initial data. 

The initial value problem for the KdV equation can be analyzed by means 
of the inverse scattering transform (IST) method \cite{Z}, pioneered in \cite{Gardner,Lax}, which relates a solution of the KdV equation (\ref{1}) to the spectrum of 
the stationary Schr\"{o}dinger equation
\begin{equation}
\label{8}
\mathcal{L} v=\lambda v, \qquad 
\mathcal{L} := -\frac{\partial^2}{\partial x^2}-u
\end{equation}
and the time-evolution equation 
\begin{equation}
\label{vt}
\frac{\partial v}{\partial t} = \mathcal{M} v, \qquad 
\mathcal{M} := -3u_x-6u\frac{\partial}{\partial x}-4\frac{\partial^3}{\partial x^3}.
\end{equation} 
The compatibility condition for the time-independent spectral parameter $\lambda$ yields the KdV equation (\ref{1}) for $u = u(t,x)$.

The IST method is usually applied on the infinite line for the initial data that decay to zero sufficiently fast at infinity. In this case, the time evolution of the KdV equation (\ref{1}) from arbitrary initial data leads to a generation of finitely many interacting solitons and the dispersive waves \cite{Deift}.  Solitons correspond to isolated eigenvalues of the discrete spectrum of the stationary Schr\"{o}dinger equation (\ref{8}) and the dispersive waves correspond to the continuous spectrum. 

For the step boundary conditions (\ref{bc}), the dynamics of the KdV 
equation (\ref{1}) are more interesting. In addition to the RW generated by the step boundary conditions, a finite number of solitary waves 
can appear from bumps in the initial data. Depending on the amplitude of these bumps, they either evolve into large-amplitude solitary waves propagating over the RW background with a constant speed or into small-amplitude solitary waves trapped by the RW  \cite{Maiden}. 

The spectrum of the stationary Schr\"{o}dinger equation (\ref{8}) for the step-like boundary conditions (\ref{bc}) was analyzed in \cite{AbCole}, where it was shown that the transmitted soliton corresponds to an isolated real eigenvalue. In regards to the trapped soliton, it was related to the so-called ``pseuso-embedded" eigenvalue located near a specific point inside the continuous spectrum which does not correspond to a true embedded eigenvalue with exponentially decaying eigenfunctions. Details of where these ``pseudo-embedded" eigenvalues are located were not given.

The rigorous IST method was applied to the KdV equation with the step-like boundary conditions in \cite{Egorova1,Egorova2}. Compared to (\ref{bc}) for the RW, the case of DSW was considered with zero boundary conditions as $x \to +\infty$. $N$ solitons scatter towards $+\infty$ as the regular KdV solitons with appropriately chosen phase shifts. The case of trapped solitons did not appear in the IST formalism. 

Here we will analyze the two scenarios of transmitted or trapped solitary waves similar to  \cite{AbCole} but in more detail. Our principal results can be summarized as follows. \\

\begin{enumerate}
	\item The transmitted solitary wave can be generated by using the Darboux transformation of the KdV equation. The Darboux transformation determines the different spatial decay rates of the solitary wave both for $x \to -\infty$ and $x \to +\infty$ by the location of an isolated real eigenvalue of the stationary Schr\"{o}dinger equation (\ref{8}). \\
	
	\item The trapped solitary wave does not actually exist. The initial condition with a ``pseudo--embedded" eigenvalue is associated with resonant poles of the stationary Schr\"{o}dinger equation which are located off the real axis and correspond to spatially decaying eigenfunctions at one infinity and growing at the other infinity.  \\
	
	\item By using numerical experiments, we show that the asymptotic amplitude of the transmitted solitary wave is determined by the initial amplitude, whereas the amplitude of the ``trapped solitary wave" decays to the amplitude of the background so that the soliton becomes invisible from the RW background for longer times.  \\
\end{enumerate}

{\bf Organization of the paper.} Section \ref{sec-2} reviews the scattering data and their time evolution for the class of solutions satisfying the boundary conditions (\ref{bc}). Section \ref{sec-3} contains examples of the initial data as the step function and a solitary wave on the step function. 
Section \ref{sec-4} gives a construction of a transmitted solitary wave on the RW background via Darboux transformation. Section \ref{sec-5} 
describes numerical simulations which illustrate that a trapped solitary wave disappears as time evolves. Section \ref{sec-6} gives a summary of our findings 
and lists open questions. \\

{\bf Notations.} We denote the Heaviside step function by $H$. The square root function $\sqrt{z}$ for $z \in \mathbb{C}$ is defined according to the principal branch such that ${\rm Arg}(\sqrt{z}) \in [0,\pi)$ for every $z \in \mathbb{C}$ with ${\rm Arg}(z) \in [0,2\pi)$. \\

{\bf Acknowledgements.} The authors thank M. A. Hoefer for guidance during the project as well as G. El and A. Rybkin for useful suggestions.
The project is supported by the RNF grant 19-12-00253.

\section{Direct scattering transform and the time evolution}
\label{sec-2}

Here we review the spectral data and their time evolution in the solutions 
of the linear equations (\ref{8}) and (\ref{vt}) for the potential $u = u(t,x)$
satisfying the boundary conditions (\ref{bc}). We assume that 
$u(t,x) \to c^2 H(x)$ as $|x| \to \infty$ sufficiently fast so that all formal expressions can be rigorously justified with Levinson's theorem for differential equations with variable coefficients which are the $L^1$ integrable perturbations of the constant coefficients. 

The linear equation (\ref{vt}) can be rewritten in the form 
\begin{equation}\label{newt}
\frac{\partial v}{\partial t} =  (4\lambda - 2u) \frac{\partial v}{\partial x} + (u_x + \gamma) v,
\end{equation} 
where we have used $v_{xxx} = - (u + \lambda) v_x - u_x v$ from 
the spectral equation (\ref{8}) and have added the parameter $\gamma$ by the transformation $v \mapsto v e^{-\gamma t}$. 

We are looking for the spatially bounded non-zero 
solutions $v = v(t,x)$. Existence of such solutions 
depend on the values of the spectral parameter $\lambda$ 
and should be performed separately in three regions: 
(1) $\lambda \in (0,\infty)$, (2) $\lambda \in (-c^2,0)$, and (3) 
$\lambda \in (-\infty,-c^2)$. The border cases $\lambda = 0$ 
and $\lambda = -c^2$ can also be included in the consideration 
but will be ignored to keep the presentation concise. \\

{\bf Case $\lambda \in (0,\infty)$.} We parameterize positive $\lambda$ as $\lambda=k^2$ with $k > 0$ and introduce 
$$
\varkappa := \sqrt{c^2+k^2}
$$
such that $\varkappa > 0$. One solution of the stationary Schr\"{o}dinger equation (\ref{8}) with $u(t,x) \to c^2 H(x)$ as $|x| \to \infty$ is given 
by $\phi(t,x;k)$ satisfying 
\begin{equation}
\label{phi-eigen}
\phi(t,x;k) \to \left\{ \begin{array}{ll} e^{-ikx}, & \quad x \to -\infty, \\
a(t;k) e^{-i\varkappa x} + b(t;k) e^{i\varkappa x}, & \quad x \to +\infty,
\end{array} \right.
\end{equation}
where the coefficients $a(t;k)$ and $b(t;k)$ are referred to as 
the scattering data. The second linearly independent eigenfunction is given by $\phi(t,x;-k)$ with the same $\varkappa$. Both solutions are bounded on $\mathbb{R}$ but not decaying to zero at infinity.

Substituting the asymptotics $\phi(t,x;k) \to e^{-ikx}$ and $u(t,x) \to 0$
as $x \to -\infty$  into (\ref{newt}), we obtain the definition of $\gamma$:
\begin{equation}\label{211}
0 = \gamma e^{-ikx} - 4 i k^3 e^{-ikx} \quad \Rightarrow \quad \gamma = 4 i k^3.
\end{equation}
Substituting the asymptotics $\phi(t,x;k) \to a(t;k) e^{-i\varkappa x} + b(t;k) e^{i\varkappa x}$ and $u(t,x) \to c^2$ as $x \to +\infty$  into (\ref{newt}) 
and using the same value of $\gamma$ from (\ref{211}),
we obtain 
\begin{align*}
& \frac{da}{dt} = i (4k^3 - 4k^2 \varkappa + 2 c^2 \varkappa) a, \\
& \frac{db}{dt} = i (4k^3 + 4k^2 \varkappa - 2 c^2 \varkappa) b,
\end{align*}
from which the exact solution is given by 
\begin{equation}
\label{data}
a(t;k)=a(0;t) e^{i(4k^2 (k-\varkappa) +2c^2\varkappa)t},  \qquad  b(t;k)=b(0;t) e^{i(4k^2 (k+\varkappa) -2c^2\varkappa)t}.
\end{equation}
Compared to the case of $c = 0$, it is no longer true that $a(t;k)$ is constant in $t$. \\

{\bf Case $\lambda \in (-c^2,0)$.} We parameterize negative $\lambda$ by $\lambda=-\mu^2$ with $\mu \in (0,c)$ and introduce
$$
\varkappa := \sqrt{c^2-\mu^2}
$$ 
such that $\varkappa > 0$. For the sake of notations, we redefine $\phi(t,x;k)$, $a(t;k)$, and $b(t;k)$ for $k = i \mu$ with $\mu > 0$ as $\phi(t,x;\mu)$, 
$a(t;\mu)$, and $b(t;\mu)$. These notations do not imply any analyticity assumptions on the eigenfunction and scattering data. The only bounded solution as $x \to -\infty$ is obtained from (\ref{phi-eigen}) with $k = i \mu$ as $\phi(t,x;\mu)$ satisfying 
\begin{equation}
\label{phi-eigen-mu}
\phi(t,x;\mu) \to \left\{ \begin{array}{ll} e^{\mu x}, & \quad x \to -\infty, \\
a(t;\mu) e^{-i\varkappa x} + b(t;\mu) e^{i\varkappa x}, & \quad x \to +\infty.
\end{array} \right.
\end{equation}
Time evolution of the scatering data is obtained from (\ref{data}) with the same change $k = i \mu$:
\begin{equation}
\label{data-mu}
a(t;\mu) = a(0;\mu) e^{(4 \mu^2 (\mu + i\varkappa) +2 i c^2\varkappa)t},  \qquad  b(t;\mu) = b(0;\mu) e^{(4 \mu^2 (\mu - i\varkappa) -2 i c^2\varkappa)t}.
\end{equation}
The second linearly independent solution $\phi(t,x;-\mu)$ is unbounded as 
$x \to -\infty$. Note that $\phi(t,x;\mu)$ decays to zero as $x \to -\infty$ 
but is not decaying as $x \to +\infty$. \\

{\bf Case $\lambda \in (-\infty,-c^2)$.} We use the same parameterization $\lambda = - \mu^2$ with $\mu > c$ and introduce 
$$
\nu := \sqrt{\mu^2 - c^2}
$$ 
such that $\nu > 0$. The only bounded solution as $x \to -\infty$ is obtained from (\ref{phi-eigen-mu}) with $\varkappa = i \nu$ so that $\phi(t,x;\mu)$ satisfies
\begin{equation*}
 \phi(t,x;\mu) \to \left\{ \begin{array}{ll}
 e^{\mu x} & \quad x \to -\infty, \\
 a(t;\mu) e^{\nu x}+b(t;\mu)e^{-\nu x} & \quad  x \to +\infty.
 \end{array} \right.
 \end{equation*}
The second linearly independent solution $\phi(t,x;-\mu)$ is unbounded as $x \to -\infty$. If $a(t;\mu) \neq 0$, then $\phi(t,x;\mu)$ is unbounded as $x \to +\infty$. However, if $a(t;\mu_0) = 0$ for some $\mu_0 \in (c,\infty)$, 
then the eigenfunction $\phi(t,x;\mu_0)$ is bounded and exponentially decaying 
as $|x| \to \infty$. The corresponding eigenfunction satisfies
\begin{equation*}
\phi(t,x;\mu_0) \to \left\{ \begin{array}{ll} e^{\mu_0 x}, & \quad x \to -\infty, \\
b_0(t) e^{- \nu_0 x}, & \quad x \to +\infty,
\end{array} \right.
\end{equation*}
where $\nu_0 := \sqrt{\mu_0^2-c^2}$ and $b_0(t)$ satisfies the time 
evolution that follows from (\ref{data-mu}):
\begin{equation*}  
 b_0(t) = b_0(0) e^{(4 \mu_0^2 (\mu_0 + \nu_0) + 2 c^2 \nu_0)t}.
\end{equation*}
Note that  $\phi(t,x;\mu_0)$ is exponentially decaying as $x \to \pm \infty$ with two different decay rates: $\mu_0$ at $-\infty$ and $\nu_0$ at $+\infty$.

\begin{remark}
We say that $\lambda_0 = -\mu_0^2$ is {\bf an isolated eigenvalue} of the stationary Schr\"{o}dinger equation (\ref{8}) if $a(t;\mu_0)  = 0$ for $\mu_0 \in (c,\infty)$ and we say that $[-c^2,\infty)$ is {\bf the continuous spectrum} of the stationary Schr\"{o}dinger equation (\ref{8}).
\end{remark}

\section{Examples of the step-like initial conditions}
\label{sec-3}

Here we solve the scattering problem for two simplest initial conditions satisfying the boundary conditions (\ref{bc}). Since the time evolution of the scattering data is not considered, we drop $t$ from the list of arguments. \\

{\bf Case of the step function $u_0(x) = c^2 H(x)$.} 
For $\lambda \in (0,\infty)$, the eigenfunction is given by (\ref{phi-eigen}), where the superposition of exponential functions hold for every $x < 0$ and $x > 0$, not just in the limits $x \to -\infty$ and $x \to +\infty$. Since $\phi$ and $\phi'$ must be continuous at $x = 0$, we derive the system of linear equations for $a(k)$ and $b(k)$:
\begin{align*}
\left\{ \begin{array}{l} 
1=a(k)+b(k),\\
-ik= i\varkappa b(k)- i \varkappa a(k). \\
\end{array} \right.
\end{align*}
The linear system admits the unique solution given by 
\begin{equation}\label{AB}
a(k)=\frac{\varkappa+k}{2\varkappa}, \qquad b(k)=\frac{\varkappa-k}{2\varkappa}.
\end{equation}
Similarly, for $\lambda \in (-c^2,0)$, the scattering data $a(\mu)$ and $b(\mu)$ are obtained from (\ref{AB}) by substituting $k = i \mu$ with $\mu > 0$:
\begin{equation}\label{AB-neg}
a(\mu)=\frac{\varkappa+i\mu}{2\varkappa}, \qquad b(\mu)=\frac{\varkappa-i\mu}{2\varkappa}.
\end{equation}
No zeros of $a(\mu)$ exists for $\lambda \in (-\infty,-c^2)$ since $a(\mu)$ is given by the same expression (\ref{AB-neg}) but with $\varkappa = i \nu$ and  $\nu + \mu = \sqrt{\mu^2-c^2} + \mu > 0$. The spectrum of the stationary 
Schr\"{o}dinger equation (\ref{8}) is purely continuous. 

\begin{remark}
	The step function can be replaced by the smooth function
	\begin{equation}
	\label{tanh-data}
	u_0(x) = \frac{1}{2} c^2 \left[ 1 + \tanh(\varepsilon x) \right], \quad \varepsilon > 0.
	\end{equation}
	Exact solutions for the scattering data $a(k)$ and $b(k)$ 
	are available in the literature \cite{MF1953}. The spectrum of the stationary Schr\"{o}dinger equation (\ref{8}) is also purely continuous. 
	We use (\ref{tanh-data}) instead of $c^2 H(x)$ in numerical experiments to reduce the numerical noise generated by the singular step function.
\end{remark}

{\bf Case of a soliton on the step function.} We consider a linear superposition of a soliton and the step function:
\begin{equation}\label{39}
u_0(x) = 2 \mu_{0}^2 \sech^2{(\mu_0(x-x_0))}+c^2H(x),
\end{equation}
where $\mu_0 > 0$ is the soliton parameter and $x_0 < 0$ is chosen to ensure that the soliton is located to the left of the step function. The direct scattering problem for the initial condition (\ref{39}) was solved in \cite{AbCole} and here we extend the solution with more details.

The spectral problem (\ref{8}) with $u = u_0$ can be solved exactly \cite{MF1953}. For $x < 0$, the exact solution for $\phi(x;k)$ satisfying 
$\phi(x;k) \to e^{-ikx}$ as $x \to -\infty$ is given by 
\begin{equation*}
\phi(x;k) = e^{-ikx} \left[ 1-\frac{i\mu_0}{k+i\mu_0}e^{\mu_0(x-x_0)}\sech{(\mu_0(x-x_0)}\right], \quad x < 0.
\end{equation*}
Similarly for $x > 0$, the exact solution for $\psi(x;k)$ satisfying 
$\psi(x;k) \to e^{i \varkappa x}$ as $x \to +\infty$ is given by
\begin{equation*}
\psi(x;k)=e^{i\varkappa x} \left[ 1-\frac{i\mu_0}{\varkappa+i\mu_0}e^{-\mu_0(x-x_0)}\sech{(\mu_0(x-x_0)}\right], \quad x > 0.
\end{equation*} 
The scattering data $a(k)$ and $b(k)$ in the representation 
(\ref{phi-eigen}) can be found from the scattering relation 
\begin{equation*}
\phi(x;k) = a(k) \overline{\psi}(x;k) + b(k) \psi(x;k), \quad x \in \mathbb{R},
\end{equation*}
where $\overline{\psi}(x;k)$ is obtained from $\psi(x;k)$ by reflection $\varkappa \mapsto -\varkappa$. Since the Wronskian $W(\psi_1,\psi_2)$ of any two solutions $\psi_1$ and $\psi_2$ of the stationary Schr\"{o}dinger equation (\ref{8}) is independent of $x$, the scattering coefficient $a(k)$ can be obtained from the formula:
\begin{equation}\label{formulaa}
a(k)=\frac{W(\phi(x;k),\psi(x;k))}{W(\bar{\psi}(x;k),\psi(x;k))}, \quad x \in \R.
\end{equation}
Since we are free to choose $x= 0$ in (\ref{formulaa}), we compute
\begin{align*}
W(\bar{\psi},\psi) |_{x=0} = & 2i\varkappa\left(1+\frac{i\mu_0e^{\mu_0x_0}\sech{(\mu_0x_0)}}{\varkappa-i\mu_0}\right)\left(1-\frac{i\mu_0e^{\mu_0x_0}\sech{(\mu_0x_0)}}{\varkappa+i\mu_0}\right)\\
& +\frac{i\mu_0^2\sech^2{(\mu_0x_0)}}{\varkappa-i\mu_0}\left(1-\frac{i\mu_0e^{\mu_0x_0}\sech{(\mu_0x_0)}}{\varkappa+i\mu_0}\right)\\
&+\frac{i\mu_0^2\sech^2{(\mu_0x_0)}}{\varkappa+i\mu_0}\left(1+\frac{i\mu_0e^{\mu_0x_0}\sech{(\mu_0x_0)}}{\varkappa-i\mu_0}\right)\\
&=2i\varkappa
\end{align*}
and
\begin{align*}
W(\phi,\psi) |_{x=0} = & i(\varkappa+k)\left(1-\frac{i\mu_0e^{\mu_0x_0}\sech{(\mu_0x_0)}}{\varkappa+i\mu_0}\right)\left(1-\frac{i\mu_0e^{\mu_0x_0}\sech{(\mu_0x_0)}}{\varkappa+i\mu_0}\right)\\
&+\frac{i\mu_0^2\sech^2{(\mu_0x_0)}}{\varkappa+i\mu_0}\left(1-\frac{i\mu_0e^{-\mu_0x_0}\sech{(\mu_0x_0)}}{k+i\mu_o}\right)\\
&+\frac{i\mu_0^2\sech^2{(\mu_0x_0)}}{k+i\mu_0}\left(1-\frac{i\mu_0e^{\mu_0x_0}\sech{(\mu_0x_0)}}{\varkappa+i\mu_0}\right)\\
&=\frac{i(\varkappa+k) (\varkappa k +\mu_0^2+i\mu_0(\varkappa-k)\tanh{(\mu_0x_0)})}{(\varkappa+i\mu_0)(k+i\mu_0)},
\end{align*}
which yields
\begin{equation}\label{final_a}
a(k)=\frac{(\varkappa+k)\left(\varkappa k +\mu_0^2+i\mu_0(\varkappa-k)\tanh{(\mu_0x_0)}\right)}{2\varkappa(\varkappa+i\mu_0)(k+i\mu_0)}.
\end{equation}
This expression coincides with (A6) in \cite{AbCole} up to notations.

Although the previous expressions were obtained for $\lambda = k^2 > 0$ with $k \in \mathbb{R}$, the scattering coefficient $a(k)$ can be continued analytically for $k \in \mathbb{C}$ with ${\rm Im}(k) \geq 0$. However, $k = ic$ is a branch point for the square root function for $\varkappa := \sqrt{c^2+k^2}$. The branch cuts can be defined at our disposal on the imaginary axis, ${\rm Re}(k) = 0$, 
for which ${\rm Im}(k)$ takes values on 
$$
\mbox{\rm either } \; [-c,c] \;\; \mbox{\rm or} \;\;
(-\infty,-c] \cup [c,\infty).
$$
We are looking for zeros of $a(k)$ for ${\rm Im}(k)>0$ 
for which $\phi(x;k) \to 0$ as $x \to -\infty$. 

\begin{itemize}
	\item If $a(k_0) = 0$ with ${\rm Im}(k_0) \in (c,\infty)$ corresponds to
$\varkappa_0 := \sqrt{c^2 + k_0^2}$ satisfying ${\rm Im}(\varkappa_0) > 0$, 
then $\phi(x;k_0) = b_0 \psi(x;k_0) \to 0$ as $x \to +\infty$. This yields {\bf the eigenvalue} $\lambda_0 := k_0^2$ of the spectral problem (\ref{8}), for which  the branch cut can be chosen for 
${\rm Re}(k) = 0$ and ${\rm Im}(k) \in [-c,c]$. In this case, the spectral theory of the Schr\"{o}dinger equation (\ref{8}) implies that ${\rm Re}(k_0) = 0$ and ${\rm Re}(\varkappa_0) = 0$. \\

\item If $a(k_0) = 0$ with ${\rm Im}(k_0) \in (0,c)$  corresponds to
$\varkappa_0 := \sqrt{c^2 + k_0^2}$ satisfying 
${\rm Re}(\varkappa_0) > 0$ and ${\rm Im}(\varkappa_0) < 0$, 
then $\phi(x;k_0) = b_0 \psi(x;k_0) \to \infty$ as $x \to +\infty$. This yields {\bf the resonant pole} $\lambda_0 := k_0^2$ of the spectral problem (\ref{8}), for which the branch cut for 
${\rm Re}(k) = 0$ and ${\rm Im}(k) \in (-\infty,-c] \cup [c,\infty)$. In this case, ${\rm Re}(k_0)$ is not generally zero.
\end{itemize}

\begin{remark}
The coefficient $b_0$ in $\phi(x;k_0) = b_0 \psi(x;k_0)$ for which $a(k_0) = 0$ can not be associated with $b(k_0)$ because the scattering coefficient $b(k)$ is not analytically continued off the real axis unlike the scattering coefficient $a(k)$.
\end{remark} 

We are now in position to analyze zeros of $a(k)$ given by (\ref{final_a}).
Since $\varkappa+k\neq0$, it follows from (\ref{final_a}) that $a(k) = 0$ if and only if $k$ is the root of the following transcendental equation 
\begin{equation}\label{zeros}
\varkappa k+\mu_0^2+i\mu_0(\varkappa-k)\tanh{(\mu_0x_0)}=0.
\end{equation}
In order to show that there exists generally a root of this equation, we can consider the limit $x_0 \to -\infty$. The algebraic equation (\ref{zeros}) is factorized in the limit $x_0 \to -\infty$ as $(\varkappa + i \mu_0) (k - i \mu_0) = 0$. Hence, there exists a simple root $k=i\mu_0$ in the limit $x_0 \to -\infty$. If $\mu_0 \in (c,\infty)$, this root corresponds to the eigenvalue $\lambda = -\mu_0^2 \in (-\infty,-c^2)$, however, if $\mu_0 \in (0,c)$, the root corresponds to the embedded eigenvalue $\lambda = -\mu_0^2 \in (-c^2,0)$ in the continuous spectrum. 

\vspace{0.2cm}
\centerline{\fbox{\parbox[cs]{0.95\textwidth}{
We claim that for $x_0 \ll -1$ an isolated 
eigenvalue $\lambda_0 \in (-\infty,-c^2)$ persists near $-\mu_0^2$ if $\mu_0  \in (c,\infty)$, whereas the embedded eigenvalue $\lambda_0 \in (-c^2,0)$ moves to a resonant pole with ${\rm Re}(k_0) < 0$ and ${\rm Im}(\varkappa_0) < 0$ if $\mu_0 \in (0,c)$. 
}}}
\vspace{0.2cm}

{\bf Isolated eigenvalue if $\mu_0 \in (c,\infty)$.} Since $\tanh(\mu_0x_0) = -1 +2e^{2\mu_0x_0}+\mathcal{O}(e^{4\mu_0x_0})$ as $x_0 \to-\infty$, the simple root of equation (\ref{zeros}) can be extended asymptotically as follows:
\begin{equation}
\label{expansion-k}
k=i\mu_0-2i\mu_0e^{2\mu_0x_0}\left(\frac{\varkappa_0-i\mu_0}{\varkappa_0+i\mu_0}\right)+\mathcal{O}(e^{4\mu_0x_0}),
\end{equation}
where $\varkappa_0 := \sqrt{c^2-\mu_0^2} = i \sqrt{\mu_0^2 - c^2}$ if $\mu_0 > c$. Therefore, $k \in i \mathbb{R}$ in the first two terms. Similarly, 
we have expansion for $\varkappa^2= c^2 + k^2$ given by 
\begin{equation}
\label{expansion-kappa}
\varkappa = \varkappa_0 \left[ 1 -\frac{2\mu_0^2}{\mu_0^2 - c^2} e^{2\mu_0x_0} \left(\frac{\varkappa_0-i\mu_0}{\varkappa_0+i\mu_0} \right) + \mathcal{O}(e^{4\mu_0x_0}) \right],
\end{equation}
so that $\varkappa\in i\mathbb{R}$ in the first two terms. In order to show that $k,\varkappa \in i \R$ persists beyond the first two terms, we substitute $k = i \mu$ and $\varkappa = i \nu$ with $\nu = \sqrt{\mu^2 - c^2}$ into (\ref{zeros}) and obtain the real-valued equation $F(\mu,\alpha) = 0$, where
\begin{equation}\label{zeros2}
F(\mu,\alpha) := \mu \sqrt{\mu^2 - c^2} -\mu_0^2+\mu_0(\sqrt{\mu^2-c^2} - \mu) \alpha, \quad \alpha := \tanh{(\mu_0x_0)}.
\end{equation}
The function $F(\mu,\alpha) : \mathbb{R}^2 \mapsto \mathbb{R}$ is a $C^1$ function near $(\mu,\alpha) = (\mu_0,-1)$ satisfying $F(\mu_0,-1) = 0$ and 
$$
\partial_{\mu} F(\mu_0,-1) = \sqrt{\mu_0^2 - c^2} + \mu_0 \neq 0.
$$
By the implicit function theorem,  there exists a simple real root $\mu \in (c,\infty)$ of 
$F(\mu,\alpha) = 0$ for every $x_0 \ll -1$ ($\alpha \approx -1$) 
such that $\mu \to \mu_0$ as $x_0 \to -\infty$ ($\alpha \to -1$). Since $k = i \mu \in i \mathbb{R}$ and $\varkappa = i \sqrt{\mu^2 - c^2} \in i \mathbb{R}$, 
the simple real root $\mu \in (c,\infty)$ determines an isolated eigenvalue 
$\lambda = - \mu^2 \in (-\infty,-c^2)$ of the spectral problem (\ref{8}). \\

{\bf Resonant pole if $\mu_0 \in (0,c)$.} Here we have $\varkappa_0 = \sqrt{c^2 - \mu_0^2} \in \mathbb{R}$ so that $k$ and $\varkappa$ in (\ref{expansion-k}) and (\ref{expansion-kappa}) 
are no longer purely imaginary. Since 
\begin{equation*}
\frac{\varkappa_0-i\mu_0}{\varkappa_0+i\mu_0} = 
\frac{(\varkappa_0 - i \mu_0)^2}{\varkappa_0^2 + \mu_0^2} 
= \frac{1}{c^2}( c^2 - 2 \mu_0^2 - 2 i \mu_0 \varkappa_0),
\end{equation*}
we obtain from (\ref{expansion-k}) and (\ref{expansion-kappa}) that
\begin{equation*}
{\rm Re}(k) = -\frac{4\mu_0^2}{c^2}\sqrt{c^2-\mu_0^2} e^{2\mu_0x_0} + \mathcal{O}(e^{4\mu_0x_0}) 
\end{equation*}
and
\begin{equation*}
{\rm Im}(\varkappa) = -\frac{4\mu_0^3}{c^2} e^{2\mu_0x_0} + \mathcal{O}(e^{4\mu_0x_0}).
\end{equation*}
Hence, ${\rm Re}(k) < 0$ and ${\rm Im}(\varkappa) < 0$ for 
the root of the complex-valued equation $F(\mu,\alpha) = 0$, which still exists for 
$x_0 \ll -1$ by the implicit function theorem.  
Therefore, the eigenfunction $\phi(x;k)$ for this root $k$ satisfies 
$\phi(x;k) \to 0$ as $x \to -\infty$ because ${\rm Im}(k) > 0$ 
but $\phi(x;k) = b_0 \psi(x;k) \to \infty$ as $x \to +\infty$ 
because ${\rm Im}(\varkappa) < 0$. Thus, this root corresponds to the resonant pole $\lambda = k^2$ with ${\rm Re}(\lambda) \in (-c^2,0)$ and ${\rm Im}(\lambda) < 0$, for which the eigenfunction $\phi(x;k)$ decays exponentially at $-\infty$ and diverges exponentially at $+\infty$.

\begin{remark}
	There exists a symmetric resonant pole $-\bar{k}$ relative to $i \mathbb{R}$ if $\varkappa = -\sqrt{c^2 + k^2}$ is defined according to the second branch of the square root function. The corresponding eigenfunction is associated 
	with the same function $\phi(x;k)$ that decays exponentially at $-\infty$ because ${\rm Im}(k) > 0$ but grows exponentially at $+\infty$ as $\phi(x;k) = b_0 \psi(x;k)$ because 
	${\rm Re}(k) > 0$, ${\rm Re}(\varkappa) < 0$, and ${\rm Im}(\varkappa) < 0$. 
\end{remark}

\begin{remark}
	It was missed in \cite{AbCole} that the "pseudo-embedded" eigenvalue near $\lambda = -\mu_0^2 \in (-c^2,0)$ splits into a pair of resonant poles. 
	There exists no embedded eigenvalues in the spectral problem (\ref{8}) if $\mu_0 \in (0,c)$.
\end{remark}

\section{A transmitted soliton via Darboux transformation}
\label{sec-4}

The Darboux transformation $u \mapsto \hat{u}$ for the KdV equation (\ref{1}) is defined as follows \cite{Matveev}. Let $u$ be a solution of the KdV equation (\ref{1}), $v_0$ be a real solution of the linear equations (\ref{8}) and (\ref{vt}) with $\lambda_0 \in \R$ such that $v_0 \neq 0$ everywhere, and $v$ be an arbitrary solution of the linear equations with arbitrary $\lambda$. Then, 
\begin{equation}\label{20}
\hat{u} := u+2\frac{\partial^2}{\partial x^2 } \log(v_0)
\end{equation}
is a new solution of the KdV equation (\ref{1}) and 
\begin{equation}\label{20a}
\hat{v} := \frac{\partial v}{\partial x} - v \frac{\partial}{\partial x} \log(v_0)
\end{equation}
is a solution of the linear equations (\ref{8}) and (\ref{vt}) with $u = \hat{u}$ for the same value of $\lambda$ as in $v$. Validity of the transformation formulas (\ref{20}) and (\ref{20a}) can be checked by direct substitutions. If $u$ is bounded, the Darboux transformation gives a bounded solution $\hat{u}$ if and only if $v_0 \neq 0$ everywhere.

\vspace{0.2cm}
\centerline{\fbox{\parbox[cs]{0.95\textwidth}{
			We claim that the Darboux transformation  
			with the step-like initial data can be used to construct 
			a transmitted soliton which corresponds to a simple isolated eigenvalue of the spectral problem (\ref{8}). 
}}}
\vspace{0.2cm}

For comparison purpose, we first construct a soliton on the zero background and then a transmitted soliton on the background of the step function $c^2 H(x)$. \\

{\bf One-soliton on the zero background.} For the trivial solution $u = 0$ of the KdV equation (\ref{1}), we can 
pick the following solution of the linear equations (\ref{8}) and (\ref{vt}) with fixed $\lambda_0 = -\mu_0^2 \in (-\infty,0)$,
\begin{equation*}
v_0(t,x) = e^{\mu_0(x-4 \mu_0^2 t-x_0)} + e^{-\mu_0(x-4 \mu_0^2 t-x_0)},
\end{equation*}
where $x_0$ is arbitrary. Substituting $v_0$ into (\ref{20}) yields the one-soliton solution 
\begin{equation}\label{one-soliton}
\hat{u}(t,x) = 2 \mu_0^2 \; {\rm sech}^2[\mu_0(x-4 \mu_0^2 t-x_0)],
\end{equation}
where $\mu_0$ determines the amplitude $2 \mu_0^2$, the width $\mu_0^{-1}$, 
and the velocity $4 \mu_0^2$ of the soliton and $x_0$ determines the initial location of the soliton. The transformation formula (\ref{20a}) for the second, linear independent solution 
\begin{equation*}
v(t,x) = e^{\mu_0(x-4 \mu_0^2 t-x_0)} - e^{-\mu_0(x-4 \mu_0^2 t-x_0)},
\end{equation*}
of the same linear equations (\ref{8}) and (\ref{vt}) with $u = 0$ and $\lambda = \lambda_0$ yields the exponentially decaying solution 
$$
\hat{v}(t,x) = 2 \mu_0 \; {\rm sech}[\mu_0(x-4 \mu_0^2 t-x_0)]
$$
of the linear equations (\ref{8}) and (\ref{vt}) with $u = \hat{u}$ and $\lambda = \lambda_0$. Hence, 
$\lambda_0 = - \mu_0^2$ is the isolated eigenvalue of the stationary Schr\"{o}dinger equation (\ref{8}) corresponding to the one-soliton solution 
(\ref{one-soliton}). 

\begin{remark}
	\label{remark-DT-cont}
Picking solutions of the linear equations (\ref{8}) and (\ref{vt}) with fixed $\lambda_0 = k_0^2 \in (0,\infty)$ does not generate bounded solutions 
of the KdV equation (\ref{1}) by the Darboux transformation. Indeed, 
a general solution is given by 
$$
v_0(t,x) = c_1 \cos(k_0 x + 4 k_0^3 t) + c_2 \sin(k_0 x + 4 k_0^3 t),
$$
where $(c_1,c_2)$ are arbitrary constants. Substituting $v_0$ into (\ref{20})  yields a new solution 
of the KdV equation (\ref{1}), 
$$
\hat{u}(t,x) = -\frac{2 k_0^2 (c_1^2 + c_2^2)}{\left[ c_1 \cos(k_0 x + 4 k_0^3 t) + c_2 \sin(k_0 x + 4 k_0^3 t) \right]^2},
$$
which is singular at countably many lines in the $(x,t)$ plane where 
$$
\tan(k_0 x + 4 k_0^3 t) = - \frac{c_1}{c_2}.
$$ 
\end{remark}

{\bf One-soliton on the initial step.} This construction is defined for the initial data at $t = 0$, hence we drop $t$ from the list of arguments similarly to Section \ref{sec-3}. For the step function $u_0(x) = c^2 H(x)$, we can pick the following solution of the stationary Schr\"{o}dinger equation (\ref{8}) with $u = u_0$ and $\lambda = -\mu_0^2 \in (-\infty,-c^2)$ with $\mu_0 \in (c,\infty)$,
\begin{equation*}
v_0(x) = \left\{ \begin{array}{ll} 
e^{\mu_0(x-x_0)} + e^{-\mu_0(x-x_0)}, & \quad x < 0, \\
c_1 e^{\nu_0 x} + c_2 e^{-\nu_0 x}, & \quad x > 0, \end{array}
 \right.
\end{equation*}
where $\nu_0 := \sqrt{\mu_0^2 - c^2} > 0$, $x_0$ is arbitrary, and $(c_1,c_2)$ are found from the continuity of $v_0$ and $v_0'$ across $x = 0$. Setting up and solving the linear system for $(c_1,c_2)$ similar to (\ref{AB}), we obtain the unique solution 
$$
\left\{
\begin{array}{l}
c_1 = \frac{\nu_0+\mu_0}{2 \nu_0} e^{-\mu_0 x_0} + \frac{\nu_0-\mu_0}{2 \nu_0} e^{\mu_0 x_0}, \\
c_2 = \frac{\nu_0-\mu_0}{2 \nu_0} e^{-\mu_0 x_0} + \frac{\nu_0+\mu_0}{2 \nu_0} e^{\mu_0 x_0}, 
\end{array}  \right.
$$
Substituting $v_0$ into (\ref{20}) yields the initial condition, 
where one soliton is superposed to the step function:
\begin{equation}
\label{22a}
\hat{u}_0(x) =  2 \mu_0^2 \; {\rm sech}^2[\mu_0(x-x_0)], \quad x < 0
\end{equation}
and
\begin{equation}
\label{22b}
\hat{u}_0(x) = c^2 + 4 \nu_0^2 \frac{\nu_0^2 + \mu_0^2 + (\nu_0^2 - \mu_0^2) \cosh(2\mu_0 x_0)}{\left[ (\nu_0+\mu_0) \cosh(\nu_0 x - \mu_0 x_0) + (\nu_0 - \mu_0) \cosh(\nu_0 x + \mu_0 x_0)\right]^2}, 
\end{equation}
for $x > 0$. The denominator of (\ref{22b}) is strictly positive for every $x_0 \in (-\infty,x_*)$, where $x_*$ is the unique positive root of the transcendental equation 
$$
\cosh(2\mu_0 x_0) = \frac{\mu_0 + \nu_0}{\mu_0 - \nu_0} > 1.
$$
If $x_0 = x_*$, the expression (\ref{22b}) is singular at $x = \mu_0 x_0/\nu_0$ 
and if $x_0 \in (x_*,\infty)$, there exist two singularities of (\ref{22b}) on $(0,\infty)$ before and after the value $x = \mu_0 x_0/\nu_0$. The transmitted soliton corresponds to the value of $x_0$ inside $(-\infty,x_*)$.

\begin{remark}
The one-soliton $\hat{u}_0$ decays differently as $x\to-\infty$ and as $x \to +\infty$, according to (\ref{22a}) and (\ref{22b}). The decay rate $\mu_0$ at $-\infty$ corresponds to the zero boundary condition, whereas the decay rate $\nu_0 = \sqrt{\mu_0^2-c^2}$ at $+\infty$ corresponds to the nonzero boundary condition $c^2$. Due to this discrepancy, the one-soliton obtained by the Darboux transformation is different from the initial condition (\ref{39}) which has the same decay rate $\mu_0$ at $\pm \infty$. In addition, the former is related to 
the isolated eigenvalue $\lambda_0 = -\mu_0^2$, whereas the latter 
is related to the isolated eigenvalue $\lambda = k_0^2$ with $k_0 = i \mu_0 + \mathcal{O}(e^{2\mu_0 x_0})$ given by (\ref{expansion-k}).
\end{remark}


\begin{remark}
The one-soliton on the initial step for $\lambda = -\mu_0^2 \in (-\infty,-c^2)$ 
corresponds to the transmitted soliton which overtakes the RW in the time dynamics of the KdV equation (\ref{1}). The corresponding time-dependent solution can be constructed from the Darboux transformation with the RW solution $u = u(t,x)$. Unfortunately, solutions $v_0(t,x)$ of the linear equations 
(\ref{8}) and (\ref{vt}) with $\lambda = \lambda_0$ are not explicit 
for the RW and therefore, it is hard to prove that 
the Darboux transformation is nonsingular with $v_0 \neq 0$ everywhere. Construction of such solutions for $t > 0$ is an interesting open problem for further studies. 
\end{remark}

\begin{remark}
	If $\lambda = -\mu_0^2 \in (-c^2,0)$ with $\mu_0 \in (0,c)$, solution of the stationary Schr\"{o}dinger equation (\ref{8}) is a bounded and oscillatory function for $x > 0$. Darboux transformation (\ref{20}) generates unbounded solution $\hat{u}_0(x)$ at a countable set of points for $x > 0$ similarly to Remark \ref{remark-DT-cont}. No trapped soliton can be constructed by the Darboux transformation for $\lambda \in (-c^2,0)$ because no embedded eigenvalues with spatially decaying eigenfunctions exist. 
\end{remark}

\begin{remark}
	If $\lambda = k_0^2 \in (0,\infty)$, then solutions of the stationary Schr\"{o}dinger equation (\ref{8}) are bounded and oscillatory both for $x < 0$ and $x > 0$. Darboux transformation (\ref{20}) generates unbounded solution $\hat{u}_0(x)$ at countable sets of points both for $x < 0$ and $x > 0$.
\end{remark}

\section{Time dynamics of one-solitons on the RW background}
\label{sec-5}

Here we perform the time-dependent computations of the KdV equation 
(\ref{1}). We utilize here the finite-difference method 
introduced by N. Zabusky and M. Kruskal in \cite{ZK} 
for numerical solution of the KdV equation (\ref{1}). 

Let $u_n^m$ be a numerical approximation of $u(t_m,x_n)$ on the equally spaced grid $\{ x_n \}$ with the equally space times $\{t_m\}$. The time step is $\tau$ and the spatial step size is $h$. The two-point numerical method in \cite{ZK} 
is given by 
\begin{equation*}
u_n^{m+1} = u_n^{m-1} - \frac{2 \tau}{h} (u_{n+1}^m+u_n^m + u_{n-1}^m) (u_{n+1}^m-u_{n-1}^m) - \frac{\tau}{h^3} (u_{n+2}^m - 2 u_{n+1}^m + 2 u_{n-1}^m - u_{n-2}^m).
\end{equation*}
The first step is performed separately with the Euler method 
\begin{equation*}
u_n^{1} = u_n^{0} - \frac{\tau}{h} (u_{n+1}^0+u_n^0 + u_{n-1}^0) (u_{n+1}^0-u_{n-1}^0) - \frac{\tau}{2 h^3} (u_{n+2}^0 - 2 u_{n+1}^0 + 2 u_{n-1}^0 - u_{n-2}^0).
\end{equation*}
The finite-difference method is stable if $\tau < \frac{2}{3 \sqrt{3}} h^3$ 
for small $h$ \cite{ZK}. To avoid oscillations of solutions due to the step background, we use the smooth function (\ref{tanh-data}) superposed with the soliton (\ref{39}) so that the initial data is 
\begin{equation}
\label{init-data}
u_0(x) = 2 \mu_0^2 {\rm sech}^2(\mu_0(x-x_0)) + \frac{1}{2} c^2 \left[ 1 + \tanh(\varepsilon x) \right], 
\end{equation}
where $x_0 < 0$ and $\varepsilon = 1$. \\

\begin{figure}[htb!]
	\centering
	\begin{subfigure}[b]{0.3\linewidth}
		\includegraphics[width=\linewidth]{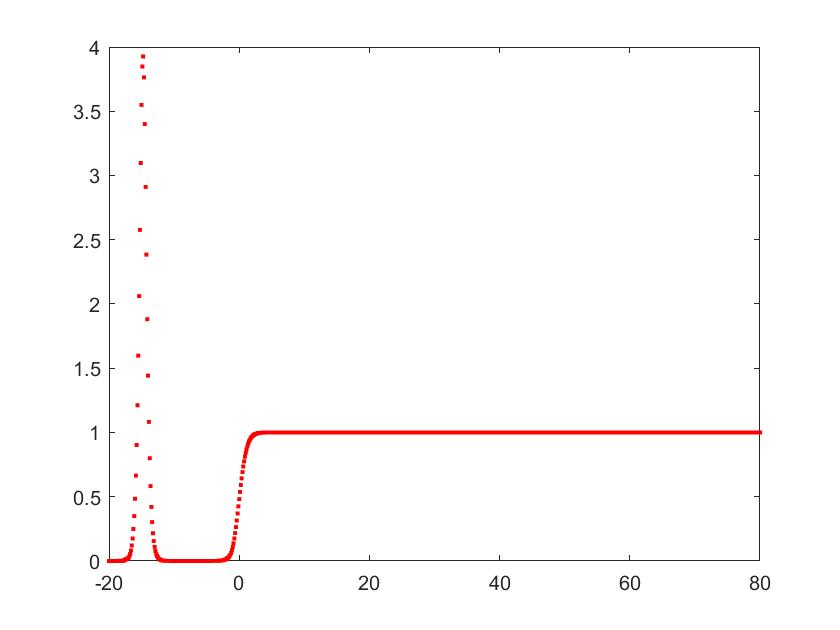}
		\caption{}
	\end{subfigure}
	\begin{subfigure}[b]{0.3\linewidth}
		\includegraphics[width=\linewidth]{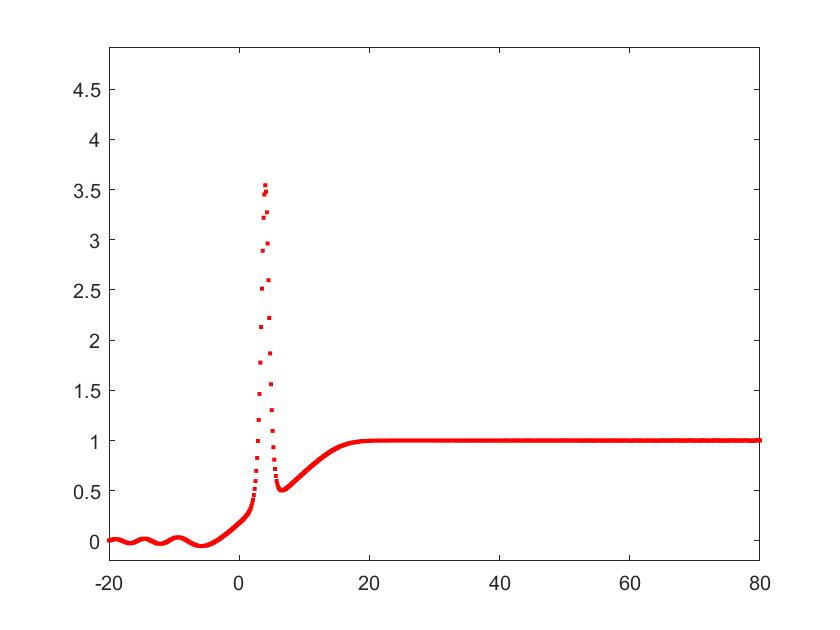}
		\caption{}
	\end{subfigure}
	\begin{subfigure}[b]{0.3\linewidth}
		\includegraphics[width=\linewidth]{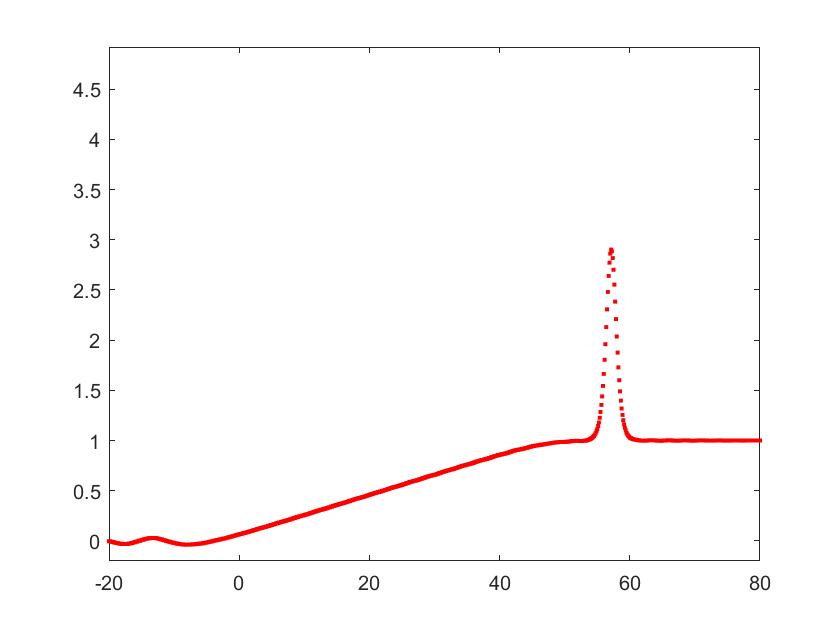}
		\caption{}
	\end{subfigure}
	\caption{The time evolution of a transmitted soliton for $\mu=1.4$, 
		$c=1$, $x_0=-15$, and $\varepsilon = 1$ at $t=0$ (left), $t=4$ (middle), and $t=8$ (right).}
	\label{fig-transmitted}
\end{figure}

{\bf Outcomes of numerical computations.} Evolution 
of the KdV equation (\ref{1}) with the initial data (\ref{init-data}) depends on the amplitude of $2\mu_0^2$ of the solitary wave to the left of the step-like background. Figure \ref{fig-transmitted} shows three snapshots 
of the evolution with $\mu_0 = 1.4 > c = 1$. A travelling solitary wave with a sufficiently large amplitude reaches and overtakes the top of the RW formed from the step-like background. This corresponds to dynamics of the transmitted soliton. 

\begin{figure}[htb!]
	\centering
	\begin{subfigure}[b]{0.3\linewidth}
		\includegraphics[width=\linewidth]{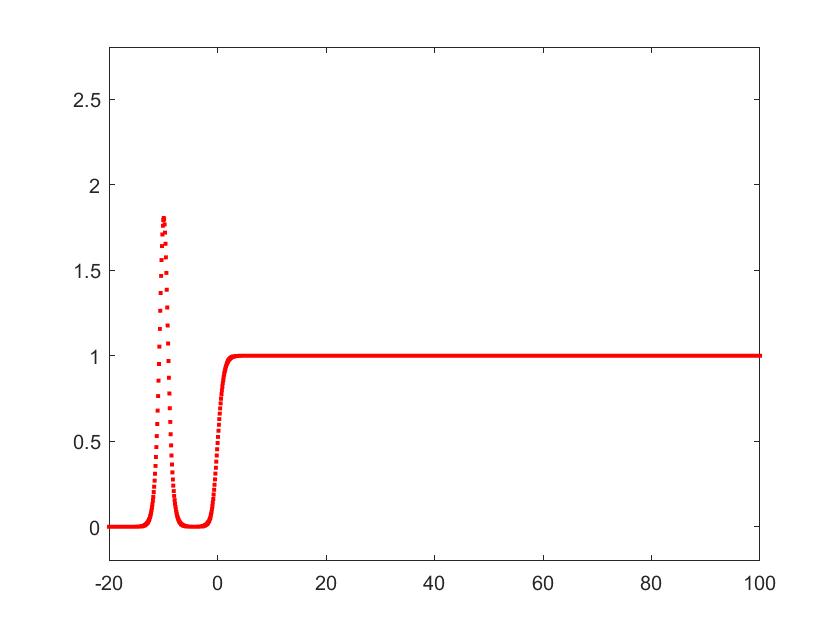}
		\caption{}
	\end{subfigure}
	\begin{subfigure}[b]{0.3\linewidth}
		\includegraphics[width=\linewidth]{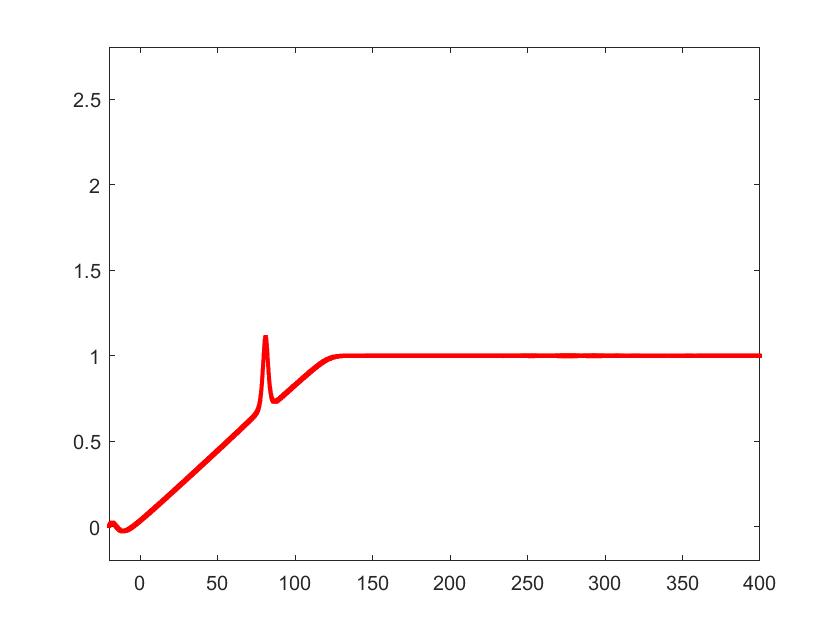}
		\caption{}
	\end{subfigure}
	\begin{subfigure}[b]{0.3\linewidth}
		\includegraphics[width=\linewidth]{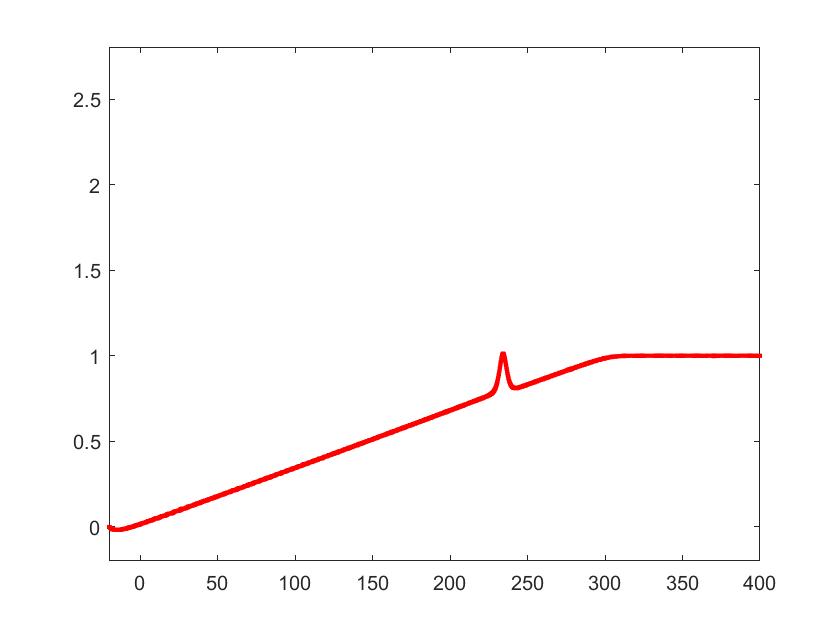}
		\caption{}
	\end{subfigure}
	\caption{The time evolution of a trapped soliton for $\mu=0.95$, $c=1$, 
		$x_0=-10$, and $\varepsilon = 1$ at $t=0$ (left) $t=20$ (middle), and $t=40$ (right).}
	\label{fig-trapped}
\end{figure}

Figure \ref{fig-trapped} shows three snapshots 
of the evolution with $\mu_0 = 0.95 < c = 1$. A travelling solitary wave with a sufficiently small amplitude becomes trapped by the RW and does not reach its top. This corresponds to dynamics of the trapped soliton. 

\begin{figure}[htb!]
	\centering
	\begin{subfigure}[b]{0.45\linewidth}
		\includegraphics[width=\linewidth]{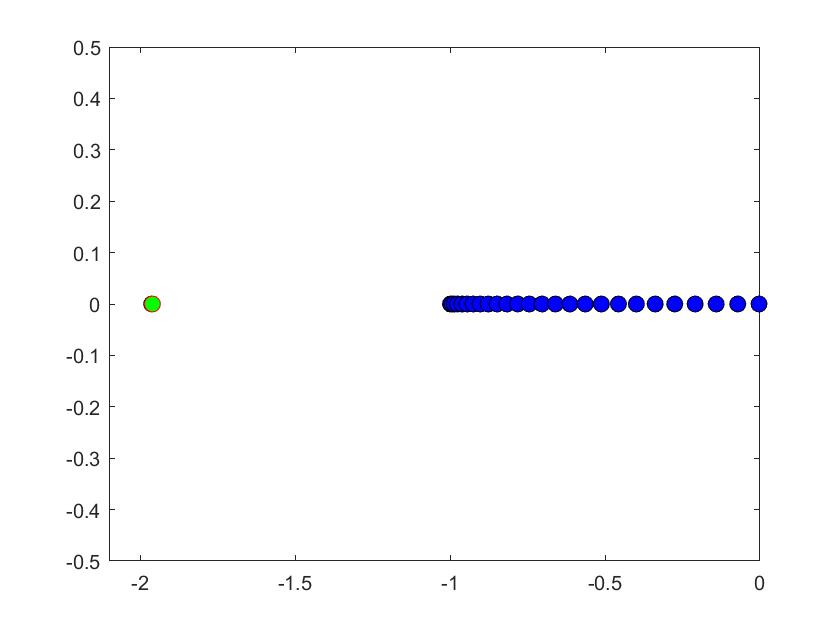}
		\caption{}
	\end{subfigure}
	\begin{subfigure}[b]{0.45\linewidth}
		\includegraphics[width=\linewidth]{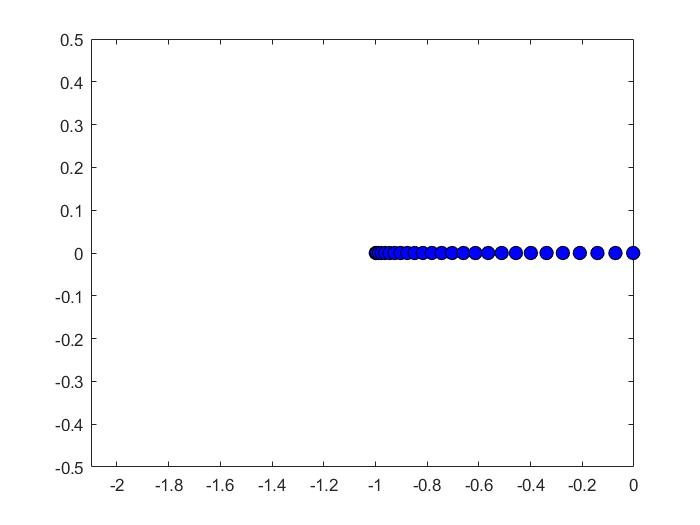}
		\caption{}
	\end{subfigure}
	\caption{Spectrum of the stationary Schr\"odinger equation (\ref{8}) for $\mu = 1.4$ (left) and $\mu = 0.95$ (right) with the potential $u_0$ given by (\ref{init-data}).}
	\label{fig:isolated}
\end{figure}

The different behavior on Figures \ref{fig-transmitted} and \ref{fig-trapped} 
is related to the different spectrum of the stationary Schr\"odinger equation (\ref{8}) shown on Figure \ref{fig:isolated} by using the second-order central-difference approximation. The left panel displays the spectrum 
for $\mu = 1.4$. The isolated eigenvalue is superimposed with the 
approximation obtained from the numerically computed root of the function (\ref{zeros2}). The difference between the two approximation 
is not visible on the scale of the figure, it is of the order of $\mathcal{O}(10^{-3})$. The right panel shows the spectrum for $\mu = 0.95$,
for which no isolated or embedded eigenvalues exist.

{\bf Data analysis.} We will elaborate the numerical criterion 
to show that the trapped soliton disappears in the long-time dynamics of the RW. In other words, the solitary wave does not appear to be a distinct soliton on the RW background but is instead completely absorbed by the RW.

Let $a^2$ be the constant background (which may change in time). The solitary wave on the constant background is obtained from the soliton on the zero background (\ref{one-soliton}) with the Galilean transformation:
\begin{equation}
\label{soliton-background}
u(t,x) = a^2 + 2 \nu_0^2 {\rm sech}^2[\nu_0(x-4 \nu_0^2 t - 6 a^2 t-x_0)],
\end{equation}
where $\nu_0 > 0$ is the soliton parameter. As follows from the construction of one-soliton on the constant background with the Darboux transformation, see 
expressions (\ref{22a}) and (\ref{22b}), $\nu_0$ is related to the fixed value $\mu_0$ (determined for $a = 0$) by $\nu_0 = \sqrt{\mu_0^2-a^2}$ as long as $a < \mu_0$. Hence the amplitude of the soliton (\ref{soliton-background}) on the constant background $a^2$ is 
$$
A = a^2 + 2 \nu_0^2 = 2 \mu_0^2 - a^2.
$$
When the solitary wave advances to the RW from the left and is strongly localized on the long scale of the RW like on Figures \ref{fig-transmitted} and \ref{fig-trapped}, the background $a^2$ is determined by the value of the RW 
at the location of the solitary wave. 

The RW can be approximated by the solution of the inviscid Burgers' equation $u_t+6uu_x=0$ starting with the piecewise linear profile 
$$
u_0(x) = \left\{ \begin{array}{ll} 0, &\quad x < -\varepsilon, \\
(2\varepsilon)^{-1} (x+\varepsilon), & \quad -\varepsilon \leq x \leq \varepsilon, \\
1, & \quad x > \varepsilon \end{array} \right. 
$$
Solving the inviscid Burgers' equation with $u(0,x) = u_0(x)$ yields
$$
u(t,x) = \left\{ \begin{array}{ll} 0, &\quad x < -\varepsilon, \\
(2\varepsilon + 6t)^{-1} (x+\varepsilon), & \quad -\varepsilon \leq x \leq \varepsilon + 6t, \\
1, & \quad x > \varepsilon + 6 t. \end{array} \right. 
$$
Location $\xi(t)$ of the solitary wave on the RW is detected numerically 
from which we determine $a^2(t) = (2 \varepsilon + 6t)^{-1} (\xi(t) + \varepsilon)$ 
as long as $\xi(t) \in [-\varepsilon,\varepsilon + 6t]$. This gives the theoretical prediction of the amplitude of the solitary wave, $A(t) = 2 \mu_0^2 - a^2(t)$, 
which can be compared to the numerical approximation of the amplitude of the solitary wave computed by the quadratic interpolation from three grid points near the maximum of $u$.

\begin{figure}[h!]
	\centering
	\begin{subfigure}[b]{0.42\linewidth}
		\includegraphics[width=\linewidth]{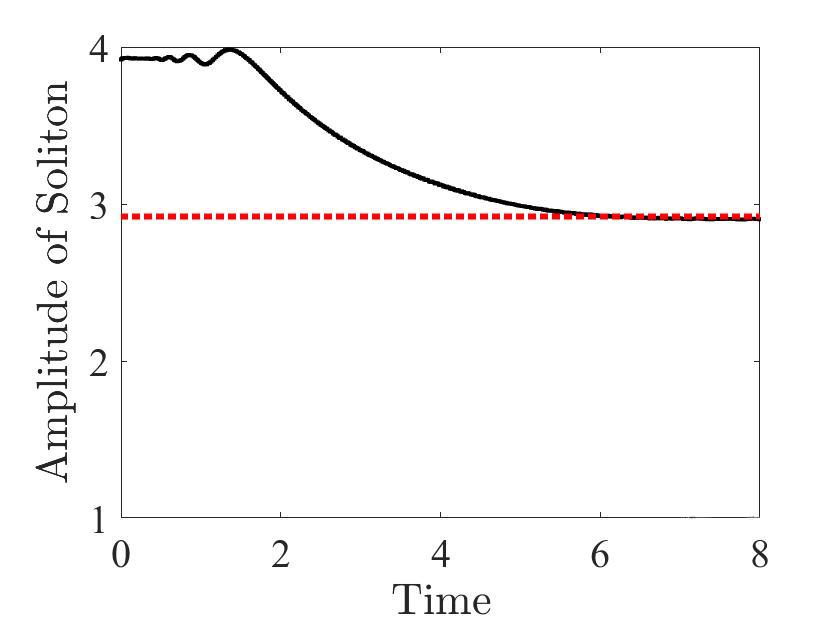}
		\caption{}
	\end{subfigure}
	\begin{subfigure}[b]{0.42\linewidth}
		\includegraphics[width=\linewidth]{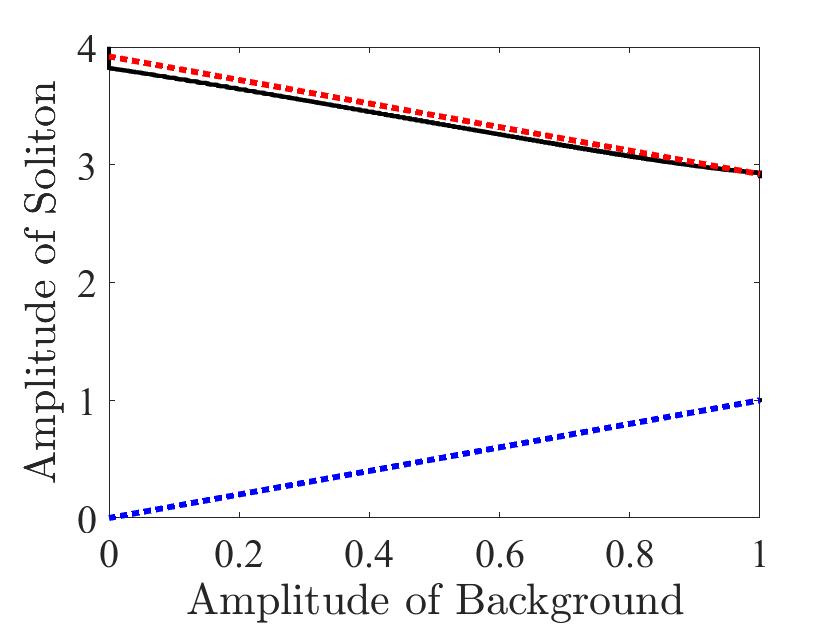}
		\caption{}
	\end{subfigure}
	\caption{Data analysis for the transmitted soliton shown in Figure \ref{fig-transmitted}: (a) Amplitude of the solitary wave versus time (black) 
		and the limiting amplitude $A_{\infty} = 2 \mu_0^2 - c^2$ (red).
		(b) Amplitude of the solitary wave versus amplitude of the RW background detected numerically (black) and theoretically (red). The blue dots show the amplitude of the RW background. }
	\label{fig:amplitude}
\end{figure}

Figure \ref{fig:amplitude} shows the numerically detected amplitude of the solitary wave versus time (left) and versus the amplitude of the RW background (right) for the transmitted soliton with $\mu_0 = 1.4 > c = 1$. The numerical approximation is shown by black dots. The red dots show the final amplitude 
$A_{\infty} = 2 \mu_0^2 - c^2$ (left) and the theoretically computed 
amplitude $A(t) = 2 \mu_0^2 - a^2(t)$ (right). It is obvious that the discrepancy between black and red dots disappear with time and that 
$A(t) \to A_{\infty}$ as $t$ evolves. The blue line on the right panel shows 
the amplitude of the background $a^2(t)$ at the location of the transmitted soliton. Since $a^2(t) \to c^2$ and $A_{\infty} > c^2$ since $\mu_0 = 1.4 > c = 1$, the black and blue lines do not meet and the soliton is transmitted over the RW background as seen in Figure \ref{fig-transmitted}.

\begin{figure}[h!]
	\centering
	\begin{subfigure}[b]{0.42\linewidth}
		\includegraphics[width=\linewidth]{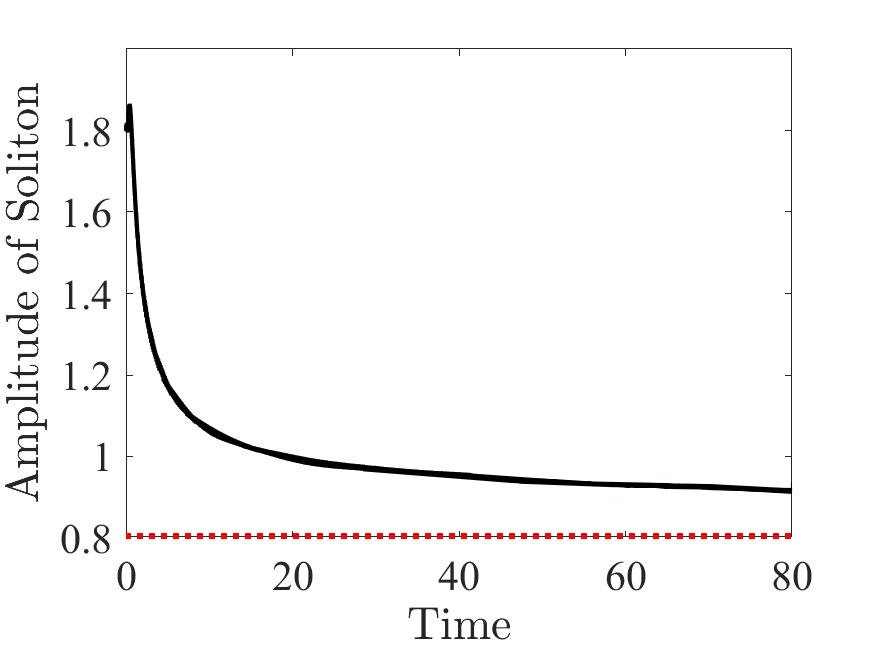}
		\caption{}
	\end{subfigure}
	\begin{subfigure}[b]{0.42\linewidth}
		\includegraphics[width=\linewidth]{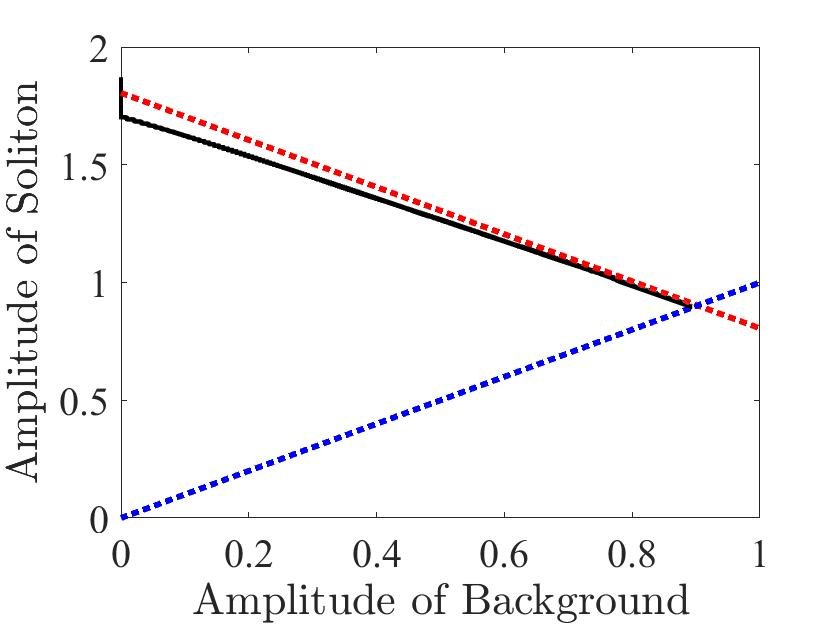}
		\caption{}
	\end{subfigure}
	\caption{The same as Figure \ref{fig:amplitude} but for the trapped soliton shown in Figure \ref{fig-trapped}.}
	\label{fig:amplitude_tr}
\end{figure}

Figure \ref{fig:amplitude_tr} shows the same quantities as Figure \ref{fig:amplitude} but for the trapped soliton with $\mu_0 = 0.95 < c = 1$. 
Since $A_{\infty} = 2 \mu_0^2 - c^2 < c^2$, the amplitude of the solitary wave never reaches the horizontal asymptote on the left panel because 
the trapped soiton dissolves inside the RW. The right panel 
shows again that the numerical approximation (black dots) is getting 
closer to the theoretical approximation of the soliton amplitude 
$A(t)$ (red dots) as $t$ evolves. However, for the trapped soliton 
the black and blue lines meet so that there exists the limiting value 
of the background $a_{\infty}^2$ such that $a(t) \to a_{\infty}$ as $t \to \infty$. The limiting value $a_{\infty}$ is found from the balance 
$2 \mu_0^2 - a_{\infty}^2 = a_{\infty}^2$ at $a_{\infty} = \mu_0$. 
Hence, the ``trapped solitary wave" is not really trapped but instead 
completely disappears inside the RW background as seen in Figure \ref{fig-trapped}.

\section{Summary}
\label{sec-6} 

We have considered the case when a solitary wave is added on the step-like initial data for the KdV equation. The step-like initial data 
evolves into a rarefactive wave (RW) whereas the solitary wave either propagates over the RW or completely disappears inside the RW. The outcome depends on 
whether there exists an isolated eigenvalue of the Schr\"{o}dinger spectral problem outside the continuous spectrum. If it exists, we can construct the transmitted soliton by using the Darboux transformation at least for $t = 0$. 
If the isolated eigenvalue does not exist, we have shown for $t = 0$ that no embedded eigenvalues exist because zeros of the transmission coefficients that correspond to the soliton data transform into complex resonant poles. 

We hope that this study will open a road for further advances on the subject 
of solitary waves propagating over the RW and DSW backgrounds. One of the important problem is to prove that the Darboux transformation remains valid 
for all values of $t \neq 0$, from which the limiting phase shifts of the transmitted solitary waves can be computed as $t \to \pm \infty$ and compared with the experimentally detected phase shifts \cite{Maiden}. Another interesting problem is to understand better how the modulated soliton theory used for data analysis in our work is justified within the Whitham modulation theory. Although the resolution formulas for $N$ solitons transmitted over the zero background have been derived in \cite{Egorova1,Egorova2}, it is interesting to see how the transformations between the two problems change these formulas to the case of the nonzero boundary conditions and how these formulas correspond to outcomes of the qualitative theory of soliton tunneling in \cite{Sande,Sprenger}.


\begin{thebibliography}{99}

\bibitem {Ablowitz} M. J. Ablowitz, {\em Nonlinear Dispersive Waves: Asymptotic Analysis and Solitons} (Cambridge University Press, Cambridge, 2011)

\bibitem{El} G. A. El, ``Korteweg-de Vries equation: solitons and undular bores", {\em Solitary waves in fluids}, Adv. Fluid Mech. {\bf 47}, 19--53 (WIT Press, Southampton, 2007).

\bibitem{ElHoffer} G. A. El and M. A. Hoefer, ``Dispersive shock waves and modulation theory", Physica D {\bf 333} (2016) 11--65.

\bibitem{Maiden}  M. D. Maiden, D. V. Anderson, A. A. Franco, G. A. El, and M. A. Hoefer, ``Solitonic dispersive hydrodynamics: Theory and observation,” Phys. Rev. Lett. {\bf 120} (2018) 144101 (5 pages).

\bibitem{Sande} K. van der Sande, G. A. El and M. A. Hoefer, ``Dynamic soliton–mean flow interaction with non-convex flux", J. Fluid Mech. {\bf 928} (2021) A21 (43 pages).

\bibitem{Sprenger} P. Sprenger, M. A. Hoefer, and G. A. El, ``Hydrodynamic optical soliton tunneling", Phys. Rev. E {\bf 97} (2018)  032218 (8 pages).

\bibitem{Congy} T. Congy, G. A. El and M. A. Hoefer, ``Interaction of linear modulated waves and unsteady dispersive hydrodynamic states with application to shallow water waves", J. Fluid Mech. {\bf 875} (2019) 1145–-1174.


\bibitem{Biondini} G. Biondini, S. Li, and D. Mantzavinos, ``Soliton trapping, transmission, and wake in modulationally unstable media, Phys. Rev. E {\bf 98} (2018) 042211 (8 pages).

\bibitem{BM} G. Biondini, S. Li, and D. Mantzavinos, ``Long-time asymptotics for the Focusing nonlinear Schr\"{o}dinger equation with nonzero boundary conditions
in the presence of a discrete spectrum", Commun. Math. Phys. {\bf 382} (2021),  1495--1577.

\bibitem{Grava} T. Grava and A. Minakov, ``On the long-Time asymptotic behavior of the modified Korteweg--de Vries equation with step-like initial Data", SIAM J. Math. Anal. {\bf 52} (2020), 5892--5993.

\bibitem{Z} S. P. Novikov, S. V. Manakov, L. P. Pitaevskii and V. E.
Zakharov,  {\em Theory of Solitons: The Inverse Scattering Method} (Consultants Bureau, New York, 1984).
	
\bibitem{Gardner} C. S. Gardner, J. M. Greene, M. D. Kruskal, and R. M. Miura, ``Method for solving the Korteweg-de Vries equation,” Phys. Rev. Lett. {\bf 19} 1095–-1097 (1967).

\bibitem{Lax} P. D. Lax, ``Integrals of nonlinear equations of evolution and solitary waves," Comm. Pure Appl. Math. {\bf 21} (1968), 467--490

\bibitem{Deift} P. Deift and E. Trubowitz, ``Inverse scattering on the line,” Commun. Pure Appl. Math. {\bf 32}  (1979) 121--251.



\bibitem{AbCole} M. J. Ablowitz, X. D. Luo, and J. T.  Cole,
``Solitons, the Korteweg--de Vries equation with step boundary values, 
and pseudo-embedded eigenvalues", J. Math. Phys. {\bf 59}  (2018), 091406 (14 pages).

\bibitem{Egorova1} I. Egorova, Z. Gladka, V. Kotlyarov, and G. Teschl, ``Long-time asymptotics for the Korteweg--de Vries equation with steplike initial data", Nonlinearity {\bf 26} (2013), 1839--1864.

\bibitem{Egorova2} I. Egorova, J. Michor, and G. Teschl, 
``Soliton asymptotics for the KdV shock waves via classical inverse scattering", 
arXiv: 2109.08423 (2021).

\bibitem{MF1953} Ph. M. Morse and H. Feshbach, {\em Methods of Theoretical Physics. Volume I} (McGraw--Hill Book Company Inc, New York, 1953).

\bibitem{Matveev} V.B. Matveev and M. A. Salle, {\em Darboux Transformations and Solitons} (Springer-Verlag, Berlin, 1991).

\bibitem{ZK} N. J. Zabusky and M. D. Kruskal, ``Interaction of “solitons” in a collisionless plasma and the recurrence of initial states", Phys. Rev. Lett. {\bf 15} 240-243 (1965).

\end{thebibliography}
\end{document}